\documentclass{eas}
\usepackage{graphicx}
\begin{document}

\newcommand{\GeVc}    {\mbox{$ {\mathrm{GeV}}/c                            $}}
\newcommand{\GeVcc}{\mbox{${\rm GeV/c^2}$}}
\newcommand{\MeVc}    {\mbox{$ {\mathrm{MeV}}/c                            $}}
\newcommand{\hetrois}    {\mbox{$ ^{3}{\mathrm{He}}                            $}}
\newcommand{\hetro}    {\mbox{$ ^{3}{\mathrm{He}}                            $}}
\newcommand{\xe}    {\mbox{$ ^{129}{\mathrm{Xe}}                            $}}
\newcommand{\ger}    {\mbox{$ ^{73}{\mathrm{Ge}}                            $}}
\newcommand{\al}    {\mbox{$ ^{27}{\mathrm{Al}}                            $}}
\newcommand{\fl}    {\mbox{$ ^{19}{\mathrm{F}}                            $}}
\newcommand{\tritium}    {\mbox{$ ^{3}{\mathrm{H}}                            $}}
\newcommand{\hequatre}    {\mbox{$ ^{4}{\mathrm{He}}                            $}}
\newcommand{\he}{$^4$He }
\newcommand{\hee}{$^4$He}
\newcommand{\het}{$^3$He }
\newcommand{\hett}{$^3$He}
\newcommand{\fe}{$^{55}$Fe }
\newcommand{\alu}{$^{27}$Al }
\newcommand{\cm}{$^{244}$Cm }
\newcommand{\iso}{C$_4$H$_{10}$ }
\newcommand{\isoo}{C$_4$H$_{10}$}
\newcommand{\nit}{N$_4$S$_3$ }
\newcommand{\neut}{$\tilde{\chi}^0$}
\newcommand{\neutt}{$\tilde{\chi}$}

\def\Journal#1#2#3#4{{#1} {\bf #2}, #3 (#4)}
\def\NCA{\em Nuovo Cimento}
\def\NIMA#1#2#3{{\rm Nucl.~Instr.~and~Meth.} {\bf{A#1}} (#2) #3}
\def\NIM#1#2#3{{\rm Nucl.~Instr.~and~Meth.} {\bf{#1}} (#2) #3}
\def\NPB{{\em Nucl. Phys.} B}
\def\PLB{{\em Phys. Lett.}  B}

\def\PRA#1#2#3{{\rm Phys. Rev.} {\bf{A#1}} (#2) #3}
\def\PRB#1#2#3{{\rm Phys. Rev.} {\bf{B#1}} (#2) #3}
\def\PRC#1#2#3{{\rm Phys. Rev.} {\bf{C#1}} (#2) #3}
\def\PRD#1#2#3{{\rm Phys. Rev.} {\bf{D#1}} (#2) #3}

\def\JHEP#1#2#3{{\rm JHEP} {\bf{#1}} (#2) #3}
\def\ZPC{{\em Z. Phys.} C}
\def\PRL#1#2#3{{\rm Phys.~Rev.~Lett.} {\bf{#1}} (#2) #3}
\def\PLB#1#2#3{{\rm Phys.~Lett.} {\bf{B#1}} (#2) #3}
\def\APP#1#2#3{{\rm Astropart.~Phys.} {\bf{B#1}} (#2) #3}
\def\APJ#1#2#3{{\rm Astrophys.~J.} {\bf{#1}} (#2) #3}
\def\APJS#1#2#3{{\rm Astrophys.~J.~Suppl.} {\bf{#1}} (#2) #3}
\def\AA#1#2#3{{\rm Astron. \& Astrophys.} {\bf{#1}} (#2) #3}
\def\JCAP#1#2#3{{\rm JCAP} {\bf{#1}} (#2) #3}

\title{MIMAC : detection of low energy recoils for Dark Matter search} 
\author{A. Trichet}\address{LPSC, Universit\'e Joseph Fourier Grenoble 1,
  CNRS/IN2P3, Institut Polytechnique de Grenoble, Grenoble, France}
\author{F. Mayet}\sameaddress{1}
\author{O. Guillaudin}\sameaddress{1}
\author{D. Santos}\sameaddress{1}

\begin{abstract}
The MIMAC project is based on a matrix of Micro Time Projection Chambers ($\mu$-TPC) for Dark Matter search, filled with He3 or CF4 and using ionization and tracks. The first measurement of the energy resolution of this $\mu$-TPC is presented as well as its low threshold.
\end{abstract}

\maketitle
%
%
Cosmological observations (\cite{wmap,archeops}) seem to point out that most of the matter in the 
Universe consists of cold non-baryonic dark matter. Nowadays, researchs are focused on Weakly 
Interactive Massive Particles (WIMPs) and especially on the lightest supersymmetric particle which 
is the neutralino in most models. Mass and elastic cross-section with ordinary matter can be computed 
in the framework of minimal SUSY models and lead to a small event rate ${\cal O} \rm (10^{-5}-1) \ day^{-1}kg^{-1}$. 
Thanks to underground laboratory and shielding, many experiments can reach this sensibility and try to observe a 
nuclear recoil through ionization, heat or light production due to WIMP elastic scaterring.\\

The MIMAC project is based on a matrix of Micro Time Projection Chambers ($\mu$-TPC), filled with $\rm ^3{He}$ or 
$\rm {CF}_4$ and using ionization and particle tracks to discriminate nuclear recoils and electrons and to 
look for an univocal directional signal due to WIMP interactions. $\rm ^3{He}$ or $\rm {CF}_4$ have significant 
advantages for Dark Matter search. They are both dominated by axial interaction which is complementary with scalar 
interaction (\cite{moulin}). Most of the WIMP events produce recoils energies below 10 keV. Moreover, the neutron capture 
is dominant process below 10 keV in $\rm ^3{He}$, with a released energy of 764 keV, leading to a
  clear signature of a neutron interaction.\\
 
\begin{figure}[thb]
\begin{center}
\includegraphics[scale=0.25,angle=270]{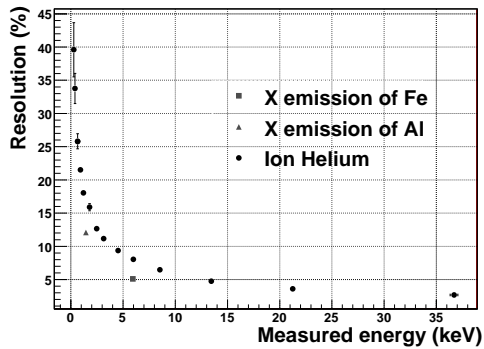}
\caption{Energy resolution ($\sigma/E$) of our µTPC for a $\rm ^4{He}$ recoil in an 95\% $\rm ^4{He}$ + 5\% Isobutan 
mixture at 700 mbar. X emission from \fe and \alu are also are also shown.}
\label{resol700}
\end{center}
\end{figure}

We will discuss of the first measurement of the low energy resolution of this $\mu$-TPC, 
developped at CEA Saclay. More details on the measurement of the quenching factor of $\rm ^4{He}$ and the experimental 
set-up may be found in (\cite{mayetlyon,MIMAC}). A good precision have been reached for all the measurements even below 10 keV. 
Currently, our team is characterizing the Micromegas detector in a gaseous mixture (95\% $\rm ^4{He}$ + 5\% Isobutan) thanks to 
an Electron Cyclotron Resonance Ion Source developped at the LPSC Grenoble coupled with a Micromegas via $1 \ \mu m $ hole with 
a differential pumping. Calibration is done with the X rays of \fe and $^{27}$Al, following energy deposition of $\alpha$ in an Al foil. 
Measurements down to 1 keV have been performed. 
The threshold of this detector in a 95\% $\rm ^4{He}$ + 5\% Isobutan mixture is around 300 eV. Fig. \ref{resol700} shows the energy
 resolution of the detector at 700 mbar. It has been shown that this energy resolution is independent of the pressure from 350 mbar 
 to 1300 mbar. Hence, it will not be problematic in the choice of the working pressure. In a co-rotational galactic model, simulations
  show that the energy resolution measured has no influence on the effective event rate and that the energy threshold corresponds to 
  an acceptable loss of 28\% of this rate. X emission from \fe and \alu are also presented. The difference between the energy
   resolution of ions X-rays and ions is possibly due to the existence of an effective mean charge for the moving $\rm ^4{He}$ ion.
 
This work, in addition to (\cite{mayetlyon,MIMAC}), shows that Quenching Factor of $\rm ^4{He}$ and energy resolution of the Micromegas have been measured precisely down to 1 keV recoil energy, thus including the range of interest for Dark Matter. In the future, our detector will be characterized thanks to a collaboration with IRSN Cadarache with a low energy neutron field in order to realize a new directional detector for Dark Matter search.
%
%
%

\end{document}